\documentclass[useAMS,usenatbib,dcolumn,usegraphicx,twocolumn,10pt]{mnarxiv}
\usepackage{times}
\usepackage[]{srcltx}
\usepackage[T1]{fontenc}
\usepackage{placeins,xspace}
\usepackage{float}
\newcommand{\eqref}[1]{(\ref{#1})}
\newcommand{\ud}[1]{\mathrm{d}#1}
\newcommand{\sqq}[1]{\left[#1\right]}
\newcommand{\vp}{\mathrm{\textit{V.p.}}}
\newcommand{\br}[1]{\left(#1\right)}
\newcommand{\mpc}{\mathrm{Mpc}}
\newcommand{\kpc}{\mathrm{kpc}}
\newcommand{\msun}{\mathrm{M}_{\sun}}

\title[]{Global disk models for galaxies NGC 1365, 6946, 7793 and UGC 6446}

\author[]{Joanna Ja{\l}ocha$^{1}$,
{{\L}ukasz Bratek$^{1}$},
{Marek Kutschera$^{1,2}$}
and {Piotr Skindzier$^{2}$}\\
$^{1}$Institute of Nuclear Physics,
Polish Academy of Sciences, Radzikowskego 152, PL-31342 Krak\'{o}w, Poland\\
$^{2}$Institute of
Physics, Jagellonian University,  Reymonta 4, PL-30059 Krak{\'o}w, Poland}

\begin{document}

\maketitle

\begin{abstract}
Spiral galaxies are studied using a simple global disc model as a means for approximate determination of mass profiles. Based on rotation curves and the amount of gas (HI+He), we find global surface mass densities consistent with measurements and compare them with B-band surface brightness profiles. As a result we obtain mass-to-light ratio profiles. We give some arguments for why our approach is reliable and sometimes better than those assuming ad hoc the presence of a massive non-baryonic dark matter halo. Using this model, we study galaxies NGC 7793, 1365, 6946 and UGC 6446. Based on a rotation curve from The HI Nearby Galaxy Survey (THINGS) we also study galaxy NGC 4536 and compare the results with those we published elsewhere for the same galaxy.\\

\medskip
\hrule
\flushleft \textbf{The definitive version is available at \\ \texttt{http://onlinelibrary.wiley.com/\\doi/10.1111/j.1365-2966.2010.16887.x/abstract}}
\medskip
\hrule
\end{abstract}


\section{Introduction}

The reliable determination of the total mass of spiral galaxies is a crucial step in the search for non-baryonic dark matter. In particular, laboratory experiments aimed at dark matter particle detection will succeed only if there is a sufficient amount of dark matter inside the disc of our Galaxy in the vicinity of the Sun.\\
\indent Present models of mass distribution in spiral galaxies assume that, in addition to the stellar disc, bulge and luminous halo, there is a massive halo of non-baryonic dark matter, most often assumed to be spherical. It dominates the total galaxy mass and also governs the dynamics of the outer parts of the stellar disc. This last feature makes it possible to test in some cases whether a spherically symmetric rather than a flattened dark matter distribution is actually present.
\\
\indent Recently, we analysed the mass distribution in several spiral galaxies
\cite{bib:22, bib:jalocha_ApJ, bib:bratek_MNRAS}, and found that for some galaxies it cannot be spherical at larger radii. Because we expect in this case that a flattened mass distribution will better approximate the gravitational potential at larger radii than a spherical one, we applied the global disc model and found that the mass distribution of luminous matter accounts for the rotation curves of the examined galaxies.\\
\indent
Furthermore, a growing number of studies, using increasingly realistic galaxy models, conclude, contrary to earlier findings, that luminous matter accounts for rotation in the internal galactic regions \citep{bib:A17,bib:A9,bib:A10,bib:A18}. There is also a possibility that the rotation of galaxies in the outermost regions could be driven by magnetic fields, not by dark matter \citep{bib:B10}.\\
\indent
These results need to be properly seen in the context of non-baryonic dark matter searches. Currently, there is strong observational evidence of non-baryonic dark matter in galaxy clusters, in particular from gravitational lensing by clusters and from colliding clusters, such as, for example, the Bullet cluster \citep{bib:25}. Furthermore, the cosmological model provides support for the non-baryonic dark matter hypothesis, albeit at a more theoretical level. On the other hand, direct searches give null results, except for DAMA/LIBRA \citep{bib:27}; however, these results are disputable and have not been confirmed by other searches.\\
\indent
A consistent picture might emerge if it is assumed that the clustering scale of non-baryonic dark matter is that of galaxy clusters. At smaller scales, non-baryonic dark matter could be dilute. In particular, its contribution to spiral galaxy masses could be much less than anticipated in spherical halo models. Moreover, it cannot be excluded that spiral galaxies may differ from each other in the abundance of dark matter -- some could be rich in dark matter, while others might be devoid of it.\\
\indent
Before presenting the results for some spiral galaxies obtained with our model based on a global disc approximation, we briefly discuss some important issues concerning rotation-curve modelling.
\\ \indent
The methods for ascertaining the mass distribution in spiral galaxies commonly use very simplified, one could even say naive, one- or many-component galaxy models that assume symmetries of mass distribution (axial, spherical), properties of rotation fields (concentric circular), and composition characteristics (constant mass-to-light ratios, etc.) that at most approximately agree with observations or sometimes even contradict them. Our model of spiral galaxies is also simplified; however, it possesses some features that, we believe, give it an advantage over other models. For example, it does not assume in advance any mass-to-light ratio nor its constancy, but a mass-to-light ratio profile is obtained as output, which seems more realistic.

In the past, the prevailing opinion was that dark matter was needed inside galaxies to account for their rotation and to stabilize their allegedly cold discs. Now, however, an increasing number of papers are modelling the same galaxies without dark matter. This is achieved by considering sufficiently complex and thus also more 'flexible' models, for example assuming wide, not thin, stellar discs etc. There is also too much freedom in choosing galaxy models, and there is no general agreement concerning which models are closer to reality.

One should try not to assume circular orbits, axial or spherical symmetry, or constant mass-to-light ratios when this is not the case for real galaxies. As pointed out by \citet{bib:C7},most spirals are not differentially rotating axisymmetric discs, with only tiny disturbances resulting from spiral arms. There are large-scale deviations from symmetry, often with very large amplitudes or even with warped HI layers \citep{bib:C7}.

The predictions of the simple galaxy models might be further modified if one reconstructed whole velocity fields rather than rotation curves, especially given that the latter are only interpretations of velocity fields obtained with additional assumptions. Going even further, different results might be obtained if, instead of using the simplified models, one used a sort of template for inferring matter distribution in real galaxies based on their velocity fields and observed qualitative characteristics such as the number of arms, the presence of a bar etc. Such templates, relating mass distribution and its qualitative shape features to simulated velocity fields, could be catalogued by elaborating a large number of galaxy-like virialized stationary systems from N-body simulations. Such an approach might offer a better way towards realistic modelling of the mass distribution in galaxies.
\subsection{Dark matter inside galaxies}
The question of how much dark matter is present in galactic interiors remains open \citep{bib:C9, bib:C8}.
\citet{bib:A9} point out that the better the data used for constraining the properties of galactic haloes and the better the quality of simulations, the more difficulties there are with dark matter models on galactic scales. In addition, \citet{bib:A10} argue that the case for cold dark matter having 'a number of stubborn problems' is weak, judged on the data from galactic scales only.\\
\indent This is especially evident from the still growing number of examples of morphologically different galaxies whose rotation curves can be accounted for without the dark matter hypothesis by using non-canonical, more realistic approaches to mass modelling. For example, even assuming a simplifying unrealistic axisymmetry and that the radial mass profile should follow that of luminosity with a constant mass-to-light ratio,
\citet{bib:A17}
were able, under the strict maximum (thick) disc hypothesis (i.e. not including a dark halo), to reproduce -- with an $I$-band mass-to-light ratio of $(2.4\pm 0.9)\, h_{75}$ and consistently with stellar population synthesis models -- the overall structure of the optical rotation curves for most galaxies in their sample of 74 spiral galaxies with various surface brightness profiles and rotation curve morphologies. They link the poorest fits with non-axisymmetric features, such as bars or strong spiral arms, that influence the brightness profiles, the determination of inclination, major-axis position angle etc. This situation might, however, be improved by considering more realistic models.\\
\indent That the interiors of spiral galaxies cannot be dominated by dark matter is also suggested by dynamical arguments. As noted by \citet{bib:A17}, if the average projected surface mass density of a dark matter halo was greater than that of the optical disc, the common instabilities would be suppressed \citep{bib:ref2} together with the lopsided modes \citep{bib:A23} appearing in many disc galaxies. In addition, as they point out, there are models that reproduce features of rotation curves within the optical discs on different scales and the models rarely require dark matter haloes \citep{bib:A19,bib:A20,bib:A21}.\\
\indent
These interesting observations of the absence of dark matter inside galaxies are best expressed by \citet{bib:A18}. They speculate, having succeeded in fitting some galaxies without dark matter, whether Occam's Razor argument might not be applicable. These authors also form less definitive statements about dark matter presence in galaxies than others. What they say is merely that dark matter presence is statistically significant.\\
\indent The presence of a dark matter component that dominates the outer mass distribution in spiral galaxies seems evident from the optical rotation curves that usually do not fall off sufficiently rapidly close to their optical edge \citep{bib:C6}, and also from extended H i rotation curves \citep{bib:C7}.\\
\indent However, in the context of what has been said above, one could conjecture that even the dark halo would disappear if the galaxy models were more realistic. In this regard one could allow for variable mass-to-light ratio profiles, take into account non-axisymmetric mass features (such as spiral arms and bars) and alter the assumed extinction models etc. Apart from that it might be fruitful to assume velocity fields more complicated than circular in the galaxy models. First of all, however, the rotation field should be known more precisely. Note that the published rotation curves to which various galaxy models are fitted are already by construction imprecise. They are obtained by reducing the two-dimensional velocity field of a galaxy (which already carries only partial information about rotation) to a single one-dimensional curve. This is performed with additional assumptions concerning the interpretation of the observed velocity field, averaging methods used, etc.
\\
\indent Concerning dark matter in the outer regions of galaxies, there is also the possibility, never taken into account in the modelling of flattened galaxies but suggested in the literature, that the rotation of slightly ionized gas at large radii may be driven by magnetic fields whose strength may be comparable with that of gravitation of the whole galaxy at these radii \citep{bib:B9,bib:B8}, helping to lower (or even eliminate) the amount of dark matter needed in the outskirts of galaxies, where rotation curves are usually flat \citep{bib:B10}.

\subsection{Dark halo and stability arguments\label{sec:stabarg}}
In the modelling of galactic optical discs, it is assumed that they are highly flattened, axisymmetric and cold. The common opinion, linked with the stability argument of \citet{bib:A2}, is that for such discs to be stable within the optical radius one necessarily needs the galaxies to be immersed in massive dark matter haloes.

This conclusion comes about as follows. As observed in numerical simulations, when random stellar motions are small compared with the streaming rotational motion in flattened galaxies, without such a halo a bar would form in one rotation period, and only afterwards would axial symmetry be regained. During this stage, approaching stationary state, the disc would heat up and could not remain cold (this would not agree with the assumption that discs are cold). Therefore, cold discs must be immersed in (usually spherical) haloes of sufficiently large mass.

There are various criteria for deciding whether a virialized rotating flattened system could be stable. \citet{bib:A2} conclude that a spherical halo with a halo-to-disc mass ratio of $1$ to $2.5$ and an initial value of the total rotational kinetic energy to total gravitational energy ratio of $0.14\pm 0.03$ renders the disc stable (of course, the halo mass beyond the outer radius of the disc has no effect and should not be taken into account). Their conclusions are supported by many computer simulations and also by linear stability analyses carried out by Toomre for a Mestel disc \citep{bib:B6}. Note, however, that for non-linear differential equations (and a hydrodynamical description of galaxies leads to such equations) it is not a priori obvious that a linear approximation appropriately describes solutions, especially given that the gravitational potential also evolves along with matter, thus also the instability modes and the associated spectrum of eigenvalues.
\citet{bib:A1} criticized the total-to-rotational energy ratio stability criterion of Ostriker \& Peebles (mentioned above), as well as various other criteria, to be of little use. Nevertheless, he points out that they all lead to similar demands on the mass distribution. As shown in \citep{bib:A1}, based on \citet{bib:C2} models, one comes to the conclusion that, for stability, the total spherical mass interior to the half-mass radius of the disc should be more than twice the disc mass in the same region. Similarly,
 \citet{bib:C3} and \citet{bib:C1} find, using their numerical models, that for the stability of differentially rotating discs the ratio of spheroidal mass to that of the disc interior to the outer radius of the disc should be approximately $2$, while for rigid discs the mass could be smaller \citep{bib:C3}. This spheroidal-to-disc mass criterion is much easier to use than the Ostriker--Peebles-like criteria in modelling galaxies, as the mass functions of the relevant galaxy components are much simpler to find.

It should be noted, however, that the assumption of cold discs is based on the belief that the local observation of a small dispersion of velocities in the solar neighbourhood in our own Galaxy can be extrapolated to the whole Galaxy, and even to all other flattened galaxies. Even if the assumption were true, there is an argument advanced by \citet{bib:A5} that the halo is not so efficient a stabilizer as a bulge alone could be. A massive halo would change the rotation field of the bar in a way not measured in real galaxies \citep{bib:A17}. As for a modest halo, it would be difficult to decide on its presence as its effect would be small \citep{bib:A5}. It is also known that the random motions of stars in the disc near the centre have a stabilizing property, as does a bulge or a thick disc. Neglecting this effect could result in an overestimation of the bulge$+$halo mass required for global stability.

\citet{bib:A1} surmises that most galaxies might have only moderate fractions of mass in the bulge$+$halo. Concerning the internal regions of the galaxy comprising the optical disc, the amount of dark matter is very poorly known \citep{bib:A17}. On the other hand, contrary to the common opinion that only spheroidal haloes can help to stabilize cold galactic discs, bulges \citep{bib:A5} or thick discs \citep{bib:ref1}  are now considered to be more efficient stabilizers \citep{bib:A15}.What is more, the presence of dark matter in large amounts, dominating the disc mass, would simply suppress the instabilities required to form the observed structures such as bars and spiral arms \citep{bib:ref2}.

Because the spherical halo mass outside the optical radius has no effect on disc stability, the facts mentioned above reverse the disc stability argument by \citet{bib:A2}, frequently used to support the need for a massive dark matter halo within the optical radius \citep{bib:A17}. Here, it is worthwhile to add that \citet{bib:B11} constructed a model disc galaxy with an almost flat rotation curve that is stable without the presence of dark matter.

\subsubsection{stability arguments and mass models}

When modelling the mass distribution in galaxies, provided that one has indeed a model that properly describes a given galaxy, one should in principle check whether the mass distribution that is found is stable. If discs were indeed cold, then for stability the spherical mass component should be sufficiently large. It seems, however, that for estimating the mass distribution in galaxies, the stability issue is not so important for astronomers, especially since the galaxy models used in this respect are very simple, one could say even naive. The majority of them assume idealized substructures such as discs and bulges with simple parametric profiles, and assume symmetries that are not observed (spherical and axial). For example, the ratio of the mass of the galaxy's spherical component to that of the disc component in the model used by  \citet{bib:blok} (THINGS) is much less than 2 for galaxies NGC 4736 and 3031. Despite the fact that a dark matter halo was included, the ratio would suggest that the obtained mass distribution is instable as it does not meet the stability criteria mentioned previously; that is, the disc-to-halo mass is too large.

There is a big 'but', however. All mass models for flattened galaxies are merely idealizations that do not take into account the various processes that make real galaxies stable. The role of the models is roughly to estimate the gravitational potential from rotation curves assuming circular orbits and axial symmetry, and so on, and then to derive the corresponding mass distribution, which is believed to some degree to approximate the actual amount of matter in real, and also stable, galaxies. For example, maximal disc models provide a substitute infinitely thin disc's surface density (that serves as a sort of a column-mass density) as a means to describe mass distribution in the whole galaxy, including the galaxy's flattened and spheroidal components. Only after a suitable de-projection of the surface density would one make estimates of the contribution from various galaxy components to the overall mass distribution. It would be therefore inappropriate to apply stability criteria to the substitute, infinitely thin discs. To address the stability problem one needs to consider more realistic, flattened mass distributions.

In real galaxies, of course, masses are distributed neither in infinitely thin discs nor in ideally spherically symmetric structures, orbits are not circular, the rotation field has a very complicated structure, and the mass distribution does not have axial symmetry -- axial symmetry is clearly broken by bars and the spiral structure. Thus it would be at least inappropriate to qualify or disqualify different models on the basis of stability arguments even if they do not contain a halo of dark matter, as the models, by construction, do not describe real galaxies and ignore many subtleties. The models are only slightly more accurate than a dimensional analysis would be, which states that the amount of electromagnetic energy radiated by a galaxy should be given by $L=\alpha D_c V_c^2/\br{G\mu}$, where $V_c$ is some characteristic velocity derived from the velocity field,  $D_c$is its radial size estimated from the spatial range of the luminous matter, $\mu$is some mass-to-light ratio derived on grounds of population analysis, and $\alpha$ is some unknown dimensionless factor related to the particular geometry of the mass distribution and the rotation field. All parameters apart from $\alpha$ can be estimated from measurements. The role of the models is to estimate the value of the parameter  $\alpha$ more precisely. For example, the estimation should be better when for an elliptical galaxy a spherical rather than a disc-like model is used, or, for a spiral galaxy, a disc model rather than a spherical one. The approximation should improve when spherical and disc-like subcomponents are distinguished; furthermore, one can consider a disc with finite width, take into account the spiral structure, and so on. However, it would be hard to treat these models seriously where the dynamics and stability of a galaxy are concerned (especially when the maximal disc model is used, which treats flat and spheroidal internal mass components on an equal footing).

Without realistic modelling, difficult problems such as structural stability cannot be solved. With realistic modelling one could try to find, for a given galaxy, a spatial rotation field and mass distribution for an N-body system -- a computer model of this galaxy -- that could be used to reconstruct at least the observed velocity field of the galaxy. Reconstruction of a published rotation curve only is not sufficient, because different groups obtain different rotation curves for the same galaxy, depending on the details of the procedures applied for processing velocity fields. Such images are also often non-symmetric, unlike simple galaxy models, and a lot of the information about galaxy kinematics that they convey is missing owing to the fact that the images are by construction a sort of projection of a three-dimensional rotational field onto the observation plane.

\subsection{Luminosity}

The mass-to-light ratio is important for studying the formation and evolution of galaxies, and thus already from this standpoint it is somewhat strange to assume that it is constant from the start, when it would be more desirable to determine it as a function of radius. Unfortunately, the theoretical mass-to-light for an individual galaxy is not well known, even when assumed constant. Its derived value depends strongly on the assumed mass function for the stellar population, on the star-formation history, on metallicity and also on extinction. It is also affected by the assumed internal extinction law, by the errors in luminosity measurements, and by the distance dimming  \citep{bib:A8}.
This ratio is generally assumed constant separately for the bulge and for the disc. Accordingly, it is tacitly assumed that the contribution of the bulge and that of the disc to the rotation curve should be controlled only by the light distribution through a single number found by best fits as if brightness profiles were to trace the overall features of rotation curves. However, the assumption of a constant mass-to-light ratio may not be correct, as it requires that extinction and population gradients across the luminous disc should be small  \citep{bib:A17}.
A constant mass-to-light ratio also seems improbable, as it would imply that the composition of a galaxy is homogenous.

Overestimating the importance of a constant mass-to-light may change considerably and in an uncontrolled way the contribution of the bulge and that of the disc to the overall rotation, especially as optical rotation curves are not featureless, while luminosity is usually fitted by simple laws (exponential, etc.). In particular, this may lead to errors in the reconstruction of the relation between the mass density and rotation of the disc component, as the rotation of a disc is a non-local functional of the density profile, sensitive to the detailed local structure of the profile. This in turn may lead to incorrect estimates of the mass profiles of other mass components. To give an example, many spirals simply do not have exponential discs or any other given by a simple analytical law, and assuming otherwise results in the large discrepancies in disc scale-lengths published by different authors for the same galaxies \citep{bib:A17}. Closely related to the non-uniqueness of the estimates of mass profiles is the so-called disc--halo degeneracy -- the uncertainty in the relative contributions of different galaxy components to the overall rotation. For unambiguous disc--halo--bulge decomposition additional constraints on mass distribution are required, for example mass-to-light ratios or the assumed mass profiles and shapes of dark haloes. In finding such constraints, helpful measurements of gravitational lensing in spiral galaxies may also be obtained \citep{bib:C5}.

The usual rotation-curve fitting methods have too many free parameters, are not unique and require additional, sometimes quite arbitrary, assumptions. For example, \citep{bib:A8} present disc--halo decompositions with several different assumptions about the stellar mass-to-light ratio for the same galaxy, or assume that only the minimum disc is present and that the observed rotation curve is attributable entirely to dark matter, or assume that the stellar mass-to-light ratio is zero(!) and take into account the contribution of the atomic gas (H i and He). In contrast, our approach does not have the arbitrariness problem, as we predict, not assume, mass-to-light ratios. It should be noted that in their more recent paper \citep{bib:blok} these authors do introduce and explore a variable stellar mass-to-light ratio derived from (J--K) colours (in the Appendix we present a comparison of these results with the predictions of our model).

As pointed out by \citet{bib:A7}, there is only limited evidence for a constant mass-to-light. What is more, the results of  \citet{bib:A6} for M33 suggest that this assumption is not valid at all: the mass-to-light ratio of the galactic disc increases approximately 5 times over 6 scale-lengths of the inner disc. Their findings also suggest that there is very little dark matter in the central regions of M33.

\section{MOTIVATION FOR THE USE OF A GLOBAL DISC MODEL}
As we have seen above, the amount of ascertained mass of dark matter is strongly model-dependent. An increasing number of papers are considering more and more complex models, compared with previous simple models of galaxies. They account for the observed rotation of internal regions of flattened galaxies without dark matter, and some of them, based on the maximal disc hypothesis, account for global rotation without dark matter. The use of models with an extended disc finds additional support from the fact that dark haloes may be very flattened rather than spherical. For flattened systems, however, gravitation becomes more complicated, and new effects that are difficult to tackle, such as the influence of external masses on internal orbits, must be taken into account that would be absent if external masses were distributed spherically symmetrically. Furthermore, stellar mass-to-light ratios need not be constant. Assuming the ratio is constant, various correlation effects between free parameters in a galaxy model may arise, such as disc--halo conspiracy, which makes rendering unique galaxy decomposition into subcomponents difficult. As we saw, this also leads to discrepancies between predictions.

Concerning the presence of dark matter in spiral galaxies, we therefore prefer a more cautious approach than just to assume the presence of a (unobserved) dark matter halo from the beginning. We propose to admit dark matter only if models composed of baryonic matter distributed in a flattened disc (of stars and gas), in the spherical bulge and in the stellar halo, fail to account properly for the dynamics of the disc measured through rotation curves. Such an approach could, we believe, significantly reduce the undesirable model dependence of the amount of cold dark matter in spiral galaxies \citep{bib:jalocha_ApJ,bib:bratek_MNRAS}.

Most models of galaxy formation predict that mass-to-light should be a declining function of distance; that is, the inner regions of a galaxy form first and contain older populations with a higher mass-to-light ratio \citep{bib:A7}. As follows from the previous section, it is incorrect to assume that the ratio is constant throughout galaxies. This is the reason why we prefer to determine the mass distribution in the disc based on measurements other than brightness profiles first, and only then determine mass-to-light ratios as a result of comparison of the obtained mass profile with the measured luminosity profile. In several galaxies we have studied so far using our approach we observe that this ratio indeed decreases while the density approaches that of hydrogen and helium known from measurements. This suggests that dark matter is not needed for some galaxies even in the outermost regions.

A conceptually simple, direct method for determining the gross mass distribution in galaxies is rotation-curve inversion. The maximal disc model is an example. Provided that the rotation curve is known globally, the corresponding mass distribution can be found through a single integral functional. The method is independent of any conventional assumptions concerning the decomposition of a galaxy into idealized subsystems with their unknown (usually correlated with each other) parameters determinable from best fits. Unfortunately, direct rotation-curve inversion is not unambiguous if the mass distribution does not have special symmetries, for example spherical symmetry.

Various ambiguities pertinent to the gravitation of flattened systems, easily hidden by the use of a particular model for the dark mass  \citep{bib:A15}, may influence the actual importance of various galaxy components. This fact does not seem to be widely appreciated. In addition, there is also ambiguity in mass-to-light ratios within optical discs, as extended rotation curves alone give no indication of where the luminous disc could end \citep{bib:B1}. Consequently, as noted in \citep{bib:A17}, there are mass models that fit rotation curves within the optical disc both with and without a significant dark component \citep{bib:B2, bib:B3}. Furthermore, the rotation of a galaxy considered to be dominated by dark matter in one model can be devoid of dark matter in another \citep{bib:jalocha_ApJ}.

For a flattened mass distribution, the local density is a non-local function of the rotation field. Vice versa, the velocity on a given orbit is affected by the gravitational forces of masses both external and internal to that orbit. For example, it would be impossible to reconstruct the dynamical mass function from the rotation of luminous matter moving in a potential well of an unseen external very flattened halo of dark matter. Such dark haloes, considerably flattened towards the stellar plane and resembling a disc more closely than a sphere \citep{bib:A15}, with axis ratios between 0.1 and 0.3, were suggested by \citet{bib:B4} and \citet{bib:B5}.
\citet{bib:bratek_MNRAS} analysed errors resulting from extrapolation of the rotation curve beyond the last measured point and gave a criterion for their estimation in global disc models. These errors are the fundamental obstacle in using disc models. The reconstructed mass density is viable only out to a distance of 60 per cent of the radius of the last measurement point. The less serious is sensitiveness of this method to the noise in rotation curve data criticized in  \citep{bib:A11}. However, as \citet{bib:A14} illustrates, this leads only to 10--20 per cent uncertainty. Therefore one needs a method to minimize such errors.

Once it is accepted that very flattened haloes of unseen matter can exist then one may lose any predictability. For example, as dark haloes are not seen, one may assume anything about them, say that outside the last measurement point there is present a flattened dark matter ring of a given arbitrarily large mass and given shape encircling the galaxy. To balance its influence on the luminous matter, one needs a correspondingly large amount of mass inside the galaxy. This in turn would make it impossible even to determine the mass-to-light ratio of the luminous matter, as the ratio would depend on the unknown mass of the flattened dark matter halo.

There is a simple criterion for deciding whether the mass distribution at larger radii in some galaxy is flattened rather than spherical. In the presence of a spherical dark matter halo, the non-spherical component of the mass distribution should be negligible. Then, only the radial component of the gravitational force is important. In this case the Keplerian mass function $G^{-1}rv_c^2(r)$ defined for rotation curve $v_c(r)$  must be a non-decreasing function of radius:  \begin{equation}\label{eq:sphcond}
\frac{\ud}{\ud{r}}\sqq{\frac{rv_c^2(r)}{G}}\geqslant0.\end{equation} In deriving this inequality, we assumed circular orbits of matter in the vicinity of the galactic plane, which is customary in the modelling of rotation curves. For spherical systems, the Keplerian mass function is identical with the true mass function. Thus, if the sphericity condition (\ref{eq:sphcond}) is not satisfied at larger radii, the true mass distribution cannot be spherical, and a spherical cold dark matter halo is thereby excluded (as far as the usual assumptions in galaxy modelling are employed), or other forms of dark matter are required forming flattened structures.

We can use the global disc model especially for galaxies with rotation curves breaking the sphericity condition at larger radii. Note that the distribution of internal masses is not very important for the overall gravitation at larger radii (the contribution from higher gravitational multipoles of internal masses decreases quickly with radius), and therefore the disc model can also be used for the inner parts, in spite of the fact that the centre may be nearly spheroidal as, after de-projection of part of the disc density onto a spheroidal component, the estimated mass of the central galaxy part is not changed significantly -- this is an approximation, a standard procedure in the maximal disc model. Because rotation curves of discs satisfying the sphericity condition are also possible \citep{bib:bratek_MNRAS}, one can try to apply the disc model also to spiral galaxies in which the rotation curves do not violate this condition.

\subsection{The model}

The disc model assumes that all the matter in a galaxy is distributed in the vicinity of the galactic equatorial plane. The model also assumes that orbits of stars are circular and that the dispersion of velocities is negligible. The other simplifying assumption is that the whole system is axisymmetric. The relation between surface mass density $\sigma(r)$ and velocity is then given by
\begin{eqnarray}\label{eq:6}
v^2(r) &=& 4Gr\vp \left(
\int\limits_{0}^{r}\sigma(\chi)\frac{\chi{}E
\br{\frac{\chi}{r}}}{r^2-\chi^2}\ud{\chi}\right.\\ &\phantom{=}&\left.
- \int\limits_{r}^{\infty}\sigma(\chi)\sqq{
\frac{\chi^2E\br{\frac{r}{\chi}}}{r\br{\chi^2-r^2}}
- \frac{K\br{\frac{r}{\chi}}}{r}}\ud{\chi}\right)\nonumber
,
\end{eqnarray}
where $K$ and $E$ are elliptic functions of the first and second kind  \citep{bib:bratek_MNRAS}. This integral is understood in the principal value sense ($\vp$), as both summands are divergent, and thus the integral must be skilfully calculated.

In practice, however, it suffices to assume that matter is located close to the galactic plane only in the outskirts of the galaxy, where the external orbits are stabilized by the gravitational potential of the galactic interior (this contribution to gravitational potential is almost spherical at large radii), while in the vicinity of the galactic interior the density can be treated as an effective density understood rather as a column density of masses projected onto the galactic plane -- a substitution for otherwise stable central spheroidal components. in internal regions. Interpreted in this way, this approximated cold disc model of an otherwise stable spiral galaxy also avoids, or at least weakens, the stability counterarguments often raised against global disc models (see also the discussion in Section \ref{sec:stabarg}).

It is seen form  equation (\ref{eq:6}) that for a flattened mass distribution of which the disc model is a limiting approximation, the rotational velocity of matter on a given orbit is dependent on masses distributed both interior and exterior to that orbit. The inverse relation is a functional of the whole rotation curve, and it reads \citep{bib:jalocha_ApJ,bib:bratek_MNRAS}
\begin{eqnarray}\label{eq:sigmamoja}\sigma(r)=
\frac{1}{\pi^2G} \vp \left[\int\limits_0^r
v^2(\chi)\biggl(\frac{K\br{\frac{\chi}{r}}}{ r\
\chi}-\frac{r}{\chi}
\frac{E\br{\frac{\chi}{r}}}{r^2-\chi^2}\biggr)\ud{\chi}
\right.\nonumber \\ \left. +  \int\limits_r^{\infty}v^2(\chi)
\frac{E\br{\frac{r}{\chi}}
}{\chi^2-r^2}\,\ud{\chi}\right].\end{eqnarray}
In the literature on this subject, for some mysterious reason, there is always used an (equivalent) expression containing derivatives of velocities that are hardly measurable with satisfactory accuracy; see, for example \citep{bib:A11}.

The crucial point now is that a rotation curve is never known globally, and thus the mass distribution even in the region where rotation is known cannot be determined from rotation only. This is a manifestation of the 'non-locality' of the gravity of flattened systems -- one cannot determine their mass functions from the observed rotation and, therefore, additional constraints on this mass distribution are required. This is particularly important in regions where measurements of rotation end, while the analogous error is negligible in the internal regions of the galaxy. This property is fatal for the reconstruction of the mass distribution inside a galaxy when outside the luminous part of this galaxy there is an external ring-like halo composed of dark matter; if it were spherical there would be no problem at all.

Having realized the unpleasant properties of flattened galaxies we devised in \citep{bib:jalocha_ApJ} a method that may help to minimize this uncertainty or remove it. One does not have to assume anything beyond the last measured point, but simply uses the available data of the mass distribution as a lower bound for the amount of matter in regions where the rotation for some reason could not be determined, but the amount of gas is still measurable. Our goal is to examine how good our model is at reproducing the mass distribution in a sample of spiral galaxies.

The role of the thin disc model is to find a global $\sigma(r)$, such that the corresponding velocity of rotation $v(r)$ defined by integral (\ref{eq:6}), best overlaps with the observed rotation curve $v_c(r)$. This task, however, is not unique, as for every R there is an infinite set of functions ?(r) that give rise to the same $R$, there is an infinite set of functions $\sigma(r)$ that give rise to the same $v(r)$ for $r<R$, even though the $\sigma(r)$'s may significantly differ from one another for $r>R$ and have different total masses. To minimize this non-uniqueness, we look for the mass distribution in flattened galaxies by iteration. A particular realization of such an iteration proposed in  \citep{bib:jalocha_ApJ} assumes that the mass distribution of gas in a given galaxy is known for radii greater than the range of the rotation curve.

We bring to the reader's attention the fact that the mass distribution found in \citep{bib:jalocha_ApJ} for galaxy NGC 4736 using this method is consistent with all measurements, and this was possible without dark matter. This result is interesting, as another model of this galaxy \citep{bib:24} predicts an almost 70 per cent abundance of dark matter and does not explain its rotation at large radii well. It is therefore natural to examine whether this galaxy is exceptional or if, perhaps, there are other spiral galaxies with lower abundances of dark matter.

\section{Results}
Below, we present the results for several galaxies obtained using the global disc model. We applied the iteration method analogous to that introduced by \citet{bib:jalocha_ApJ}.
We are aware that this is not a representative sample of galaxies. However, we think that if the rotation of even a single galaxy could be explained without dark matter this would be interesting and worthy of presentation.

Initially, we planned to examine only galaxies with rotation curves breaking the condition of a spherical mass distribution, because such galaxies are presumably flattened and, therefore, natural candidates for application of the global disc model. However, in order to see whether the model produces viable results for other galaxies, we included also galaxies NGC 6946 and UGC 6446, for which we also found good and complete data.

 Concerning the error analysis, our method by construction exactly reproduces the rotation curves, and thus also fits rotation measurements within error bars. This is the reason why we show rotation curves without error bars.

\subsection{NGC 4736}

\begin{figure}
\includegraphics[width=\columnwidth,height=2\columnwidth]{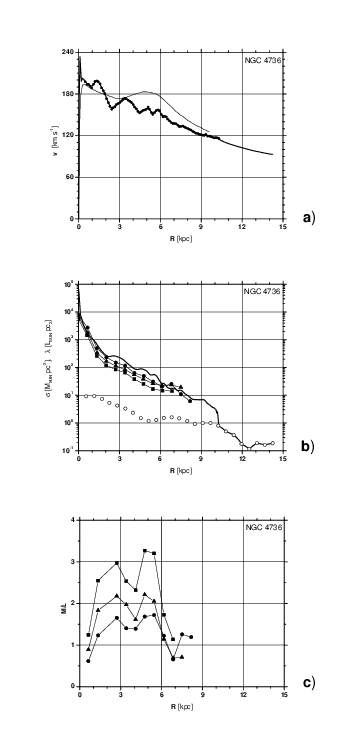}
\caption{\label{fig:4736}
Results obtained using the global disc model for galaxy NGC 4736 at $D=4.7 Mpc$.
\textbf{a)}  Rotation curve: THINGS measurements
\textit{[solid circles]}
\citep{{bib:blok}},
the model \textit{[solid line]},
and comparison with  high resolution rotation curve taken from \citep{bib:sofue}.
\textbf{b)} model global surface mass density \textit{[solid line]},
surface mass density of HI+He \textit{[open circles]}; and surface brightness: B-band \textit{[solid squares]}, V-band \textit{[solid triangles]}, I-band \textit{[solid circles]} \citep{bib:munoz}; \textbf{c)} mass-to-light ratio profile without inclusion of HI+He.
}
\end{figure}
Here, we once again consider galaxy NGC 4736, which was previously studied in \citep{bib:jalocha_ApJ}. This time, however, we use another, newer rotation curve published by \citet{bib:blok}. In Fig. \ref{fig:4736},
this new curve is compared with the one obtained by \citet{bib:sofue} that we used last time. Both the curves are shown assuming a distance of $4.7 \mpc$ different from the previous $5.1 \mpc$. Although the two curves do not overlap we come to comparable conclusions concerning the amount of non-baryonic dark matter. NGC 7436 remains a galaxy in which the halo is not necessarily required. The surface density obtained in the global disc model smoothly approaches that of gas in the outskirts of the galaxy; we also obtain low mass-to-light ratios ($1.28$ for the I-band, $1.44$ for the V-band, and $2.01$ for the B-band).
In addition, the local mass-to-light ratio decreases with radius for the largest radii. Therefore, despite the fact that the new rotation curve does not break sphericity conditions, in contrast to the curve obtained by \citet{bib:sofue}, we still consider this galaxy to contain no dark matter or that the abundance of dark matter is at most very low. In the Appendix we compare our results concerning the mass-to-light ratio profile of this galaxy with the results of \citep{bib:blok}.


\subsection{NGC 7793}
\begin{figure}
\includegraphics[width=\columnwidth,height=2\columnwidth]{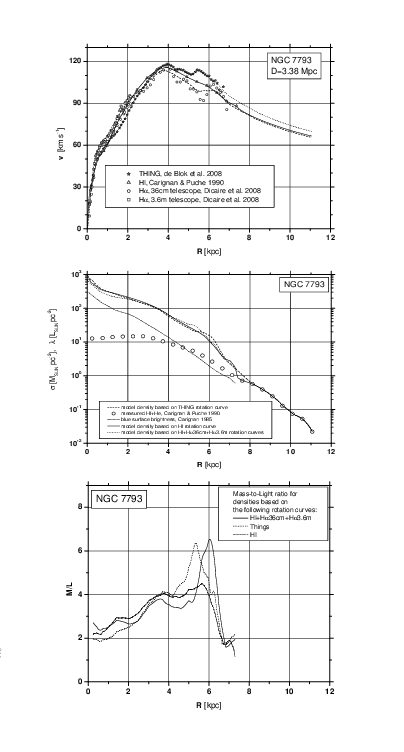}
\caption{\label{fig:7793}
\textit{Upper:} Comparison of the NGC 7793 rotation curves obtained in THINGS \citep{bib:blok}, \citep{bib:carignanpuche} and \citep{bib:dicaire}.  Symbols stand for measurements and lines for model fits. \textit{Middle:} surface mass densities obtained in global disk model compared with surface mass densities of neutral hydrogen and helium and with surface brightness in the B-band. \textit{Bottom:} Mass-to-light ratios derived based on mass densities obtained for different rotation curves.
}
\end{figure}

We take (i) the rotation curve published by \citet{bib:carignanpuche}, (ii) that by \citet{bib:blok} and (iii) three rotation curves published by
\citet{bib:dicaire}. The latter three curves were merged so as to obtain a single rotation curve. In effect we have three distinct rotation curves for the same galaxy. As seen in Fig. \ref{fig:7793} the curves are similar, which shows that the rotation measurements for this galaxy are trustworthy. The corresponding masses are also close to each other, $1-1.02 \times10^{10}\msun$. The surface mass density from the global disc model smoothly converges to that of gas at the outskirts of the galaxy for all three cases. We also obtain low values for mass-to-light ratios  ($2.93-3$ in the B-band).
We stress that the attempt of {\citet{bib:blok}}  to account for the rotation with a spherical cold dark matter halo gives very poor results. By giving up the halo, we obtain a surface mass density in agreement with the amount of gas, and low mass-to-light ratios. Therefore, we think it is reasonable to assume that the galaxy does not have a dark matter halo.

\subsection{NGC 1365}

\begin{figure}
\includegraphics[width=\columnwidth,height=2\columnwidth]{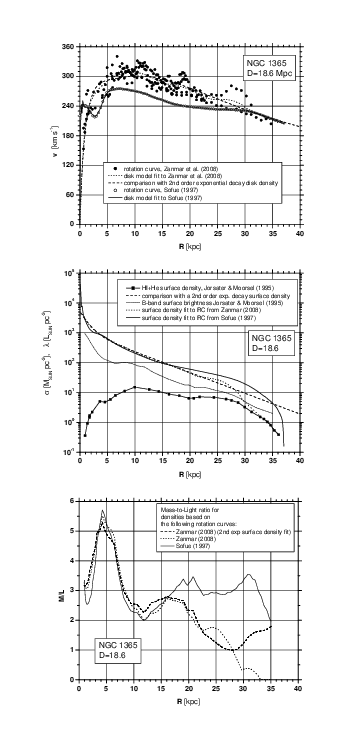}
\caption{\label{fig:1365}
\textit{Upper:} Comparison of the NGC 1365 rotation curves obtained in \citep{bib:zanmar} and \citep{bib:sofue}.  Symbols stand for measurements and lines for model fits. \textit{Middle:} surface mass densities obtained in global disk model compared with a 2nd order exponential decay fit and with surface mass densities of neutral hydrogen and helium and surface
brightness in the B-band.
\textit{Bottom:}  Mass-to-light ratios derived
based on mass densities obtained different rotation curves.
}
\end{figure}

We now consider two rotation curves, namely \citep{bib:sofue} and \citep{bib:zanmar} the latter suitably smoothed out in a way to conform with data points and error bars. We have noticed that the surface mass density corresponding to the rotation curve has an approximately exponential falloff (see Fig. \ref{fig:1365}).
It is therefore justified to assume first that the surface mass density is a superposition of two exponential profiles, in order to check whether such mass distribution accounts for the observed rotation. We find that this is indeed the case; however, the mass density becomes larger than that of gas (although quickly falling off). On the other hand, the surface density obtained with our method converges smoothly to that of gas at a place where the galaxy is still bright in the B band. This can be interpreted as that there must be luminous matter other than HI and He present. It is very interesting that for this galaxy the doubly exponential disc alone suffices to account for the rotation curve of  \citep{bib:zanmar}. Irrespective of which rotation curve is considered, the obtained mass-to-light ratio is low ($2.91$ to $3.15$ in the B band depending on which rotation curve and method is used). This ratio falls off with increasing radius in every case. Thus, galaxy NGC 1365 can be modelled without non-baryonic dark matter.

\subsection{NGC 6946}

\begin{figure}
\includegraphics[width=\columnwidth,height=2\columnwidth]{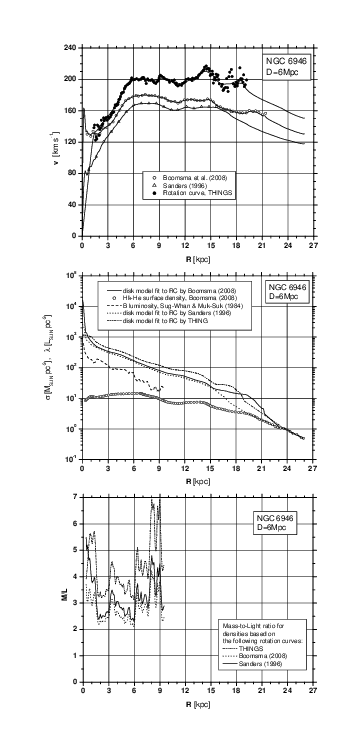}
\caption{\label{fig:6946}
\textit{Upper:} Comparison of the NGC 6946 rotation curves obtained in \citep{bib:boomsma}, \citep{bib:sanders} and THINGS \citep{bib:blok}.  Symbols stand for measurements and lines for model fits. \textit{Middle:} surface mass densities obtained in global disk model compared with surface mass densities of neutral hydrogen and helium and surface
brightness in the B-band.
\textit{Bottom:}  Mass-to-light ratios derived
based on mass densities obtained different rotation curves.
 }
\end{figure}

We take three rotation curves published by \citet{bib:sanders},
by \citet{bib:boomsma} and  \citet{bib:blok}
(the latter were smoothed out). As can be seen in Fig. \ref{fig:6946} the published rotation data differ significantly from each other; in particular, the rotation curve of  \citet{bib:blok} greatly exceeds the others. These differences result partly from the various inclination angles assumed. By using a global disc model we obtained surface densities that smoothly converge to that of gas. The mass-to-light ratio is low based on the rotation by Sanders  ($3.49$ in the B-band),
but it becomes twice as large for the rotation curve of THINGS  ($6.15$). The mass-to-light ratio for THINGS grows slightly with radius; for the other two curves it is approximately constant. Summing up, based on the available rotation data it is difficult to decide uniquely about the presence of non-baryonic dark matter in this galaxy. The result depends of which rotation curve is used.

Using our modelling method we can explain all these rotation curves. It is worthwhile to note that the obtained surface densities smoothly converge to that of the gas. However, based on the THINGS rotation curve, we obtain an mass-to-light ratio much greater than for that NGC 4736, 1365 and 7793, and similarly for the other two rotation curves.

\subsection{UGC 6446}
\begin{figure}
\includegraphics[width=\columnwidth,height=2\columnwidth]{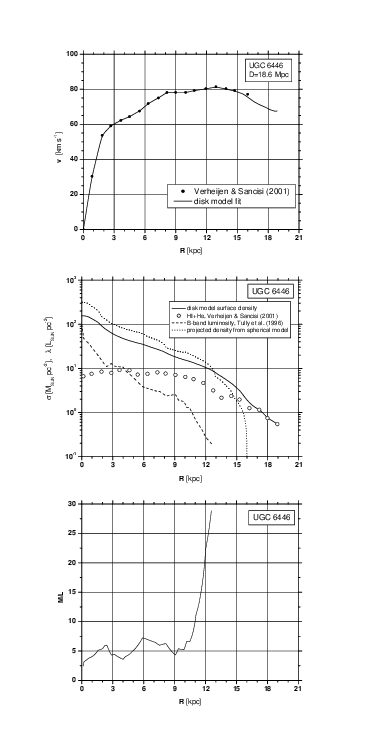}
\caption{\label{fig:6446}
\textit{Upper:} Comparison of the NGC 6446 rotation curve obtained in \citep{bib:19}.  Symbols stand for measurements and lines for model fits. \textit{Middle:} surface mass density obtained in global disk model compared with that of obtained in spherical mass model, with that of neutral hydrogen and helium and with surface
brightness in the B-band.
\textit{Bottom:}  Mass-to-light ratio derived
based on mass density obtained different rotation curves.
}
\end{figure}

The mass-to-light profile of UGC 6446 grows with radius, attaining values much higher than those for the other examined galaxies, even after subtraction of HI and He. Such a singular behaviour of the mass-to-light profile is expected at the edge of a luminous disc, where the luminosity per unit of surface tends to zero, $\lambda(r) \to 0$, while the surface mass density remains finite. Then, the local mass-to-light ratio becomes singular,
$\sigma(r)/\lambda(r) \to \infty$. After subtraction of the mass contribution of gas, the increasing mass-to-light ratio indicates that the stars at the outskirts of the luminous disc contain a possible admixture of dead stars, 'black' dwarfs, neutron stars, and stars of too low a mass for nuclear burning (so-called brown dwarfs). It is important to note that the mass density of such a dead star component needs only to be residual, such that the total mass density is infinitesimally bigger than the mass density of shining stars, that is $\sigma_*+\epsilon$  for the mass-to-light ratio significantly increases at the edge:
$(\sigma_*+\epsilon)/\lambda \to \epsilon/\lambda$ when both $\sigma_*(r)\to 0$ and $\lambda(r) \to 0$. This very small mass density, $\epsilon$, of dead stars does not show up anywhere except at the very edge, where it is strongly enhanced by small  $\lambda(r)$. The contribution of dead stars to the mass of the galaxy is thus negligible.

The following thought experiment provides a strong argument against the existence of a spherical halo around UGC 6446. Suppose that this galaxy is dominated by such a halo. Then, the mass distribution could be considered spherical with good accuracy, and thus the corresponding mass function could be calculated from  $M(r)=\frac{rv^2(r)}{G}$(and this can be done, as the rotation curve of this galaxy satisfies the sphericity condition). Then, the corresponding column mass density  $\sigma_{col.}$ can be found by projecting the volume density onto the galactic plane using the formula
\begin{eqnarray} &\mu(r):=\frac{rv^2(r)}{G};\qquad \sigma_{col.}(\rho)=
\frac{\mu(R)-\mu\br{\rho}}{2\pi{}R\sqrt{R^2-\rho^2}}+\dots&\\ &+\lim
\limits_{\epsilon\to0}\frac{1}{2\pi}{\int\limits_{\sqrt{
\epsilon^2+\rho^2}}^R\frac{2r^2-\rho^2}{r^2\br{r^2-\rho^2}^{3/2}}\sqq{\mu(r)-
\mu\br{\sqrt{\epsilon^2+\rho^2}}} },&\nonumber\end{eqnarray} where $R$ is the distance of the outermost point on the rotation curve. Note that our formula is very advantageous from a practical point of view, in contrast to the (equivalent) handbook formula, as it does not require derivatives of velocities, which are measured with very poor accuracy. We compare the column density $\sigma_{col.}$ with the total surface mass density obtained from the disc model and with its component surface mass density of gas (HI+He). The result is shown in Fig. \ref{fig:6446}.
We see that close to the 'edge', $\sigma_{col.}$ is lower than the amount of gas! We stress that the two surface mass densities seen in Fig.  \ref{fig:6446} correspond to two extreme idealizations -- in the first the galaxy is approximated by a thin disc, and in the other, by a spherically symmetric mass distribution. The realistic mass distribution must be somewhere in between, but, as is seen, closer to that in the disc model. In view of these observations, we classify UGC 6446 as a galaxy without a massive spheroidal dark halo.

\section{Summary}
We considered five galaxies, one of which was re-examined with a new rotation curve. The results are summarized in Table \ref{tab:2}. Our aim was to determine whether the galaxies could be satisfactorily modelled without non-baryonic dark matter. In this regard we checked whether the following conditions were met: the mass distribution obtained should account for the measured rotation curve; it should approximately converge towards that observed for gas (HI + He) in the outer parts of galaxies (as this is the lower bound for the amount of luminous matter); the mass-to-light ratio should be low and its profile should be a declining function of radius close to the galaxy 'edge'.

We find that these conditions are satisfied by our results for galaxies NGC 4736, 7793 and 1365 for all rotation curves. It is reasonable to state that these are galaxies with flattened, disc-like mass distributions, for which the disc model approximation works quite well. The case of galaxy NGC 6946 is not so clear. There are significant discrepancies between published rotation curves, and consequently in the obtained mass-to-light ratios. Galaxy UGC 6446 satisfies some of our criteria for the absence of dark matter; however its mass-to-light ratio is highest of all the galaxies we examined in this paper and, in addition, it grows dramatically towards the galaxy 'edge'. In another test we assumed that the galaxy was dominated by a spherical dark matter halo and, using a spherical model, obtained the resulting bulk mass density. This density, accounting for the galaxy rotation, is in this case lower than that of gas, which can be considered as a strong argument for the absence of a spherical non-baryonic dark halo. This argument does not exclude the presence of a non-spherical halo. However, the mass needed in the galaxy outskirts to account for the rotation in the disc model is practically that of the gas. It would therefore be worthwhile to consider causes other than cold dark matter for the mass-to-light ratio growth.

We have also raised the controversial issue of to what extent the very simplified galaxy models, such as the ones commonly used by astronomers and in the current paper, can be treated as models of realistic galaxies. In particular, we examined whether stability arguments can be applied to favour or reject a particular galaxy model. We stressed that the role of these idealized models is not to reconstruct the exact spatial mass distribution, but to estimate various global characteristics of galaxies such as the total mass, amount of gas, global mass-to-light ratio, dark-to-luminous mass ratio, etc. In particular, various effects, such as the influence of galactic magnetic fields on the rotation of gas in the galaxy outskirts, are normally not taken into account, and these fields may be important for accounting for flat rotation curves \citep{bib:B10}.

A global disc model is a natural approximation for flattened galaxies such as spiral ones, as far as it is not assumed in advance, sometimes deliberately, that each spiral galaxy must necessarily be immersed in an invisible halo of dark matter. Moreover, we are able to find such mass distributions that perfectly account for rotation curves, while in models assuming a massive halo these rotation curves often cannot be satisfactorily reconstructed, for example in the case of NGC 7793 in \citep{bib:blok}, or the dark halo obtained is too weak to account for stability criteria (NGC 4736 in \citep{bib:blok}) (whereas historically the role of such haloes was not only to explain flat rotation, but also to stabilize discs). Although our model is very simple, as are other galaxy models commonly used by astronomers, it is better in that it does not assume constant mass-to-light ratios but predicts the ratio as a resultant mass-to-light profile. Our results pertain only to the particular galaxies examined by us so far. From the fact that there is no need of non-baryonic dark matter in galaxies NGC 4736 and 7793, one cannot conclude that it is not needed in other spiral galaxies. To make such a claim, further studies are needed. Nonetheless, the existence of spiral galaxies without non-baryonic dark matter and of galaxies with a small abundance of such matter is interesting.

\begin{table*}
\begin{minipage}{2\columnwidth}
\begin{tabular}{@{}r@{}|@{}c@{}|@{}c@{}|@{}c}
\hline
name     &
NGC 7793 &
NGC 1365 &
NGC 6946 \\ 
\hline
incl. angle [deg] &
$53.7$ ${}^A$ &
$46$ ${}^B$   &
$30$ ${}^B$   \\
distance [Mpc] &
$3.38$ ${}^A$ &
$18.6$ ${}^L$ &
$6.0$ ${}^M$ \\
morphological type &
SAd ${}^A$ &
SBb ${}^B$ &
SAbc ${}^B$ \\
$L_{B}$ $\sqq{10^{10}L_{\odot}}$ &
$0.3$ ${}^K$ &
$9.81$ ${}^E$ &
$1.64$ ${}^H$ \\
$M_{H+He}$ $\sqq{10^{10}M_{\odot}}$ &
$0.12$ ${}^A$ &
$2.36$ ${}^E$ &
$0.97$ ${}^M$ \\
$M_{H_2}$ $\sqq{10^{10}M_{\odot}}$ &
- &
- &
$0.45$ ${}^J$ \\
\begin{tabular}{r}
rotation curve \\
$M_{Gal}$ $\sqq{10^{10}M_{\odot}}$ \\
breaking sphericity test\\
${M_{Gal}}/{L_{B}}$\\
$\frac{M_{Gal}-M_{gaz}}{L_{B}}$
\end{tabular}
&
\begin{tabular}{c|c|c}
$HI$ ${}^A$ & $HI+H_{\alpha}$ ${}^O$ & Thing ${}^P$\\
$1$ & $1.02$ & $1$\\
Y&Y&Y\\
$3.33$ & $3.4$ & $3.33$\\
$2.93$ & $3$ & $2.93$\\
\end{tabular}
&
\begin{tabular}{c|c|c}
${}^B$ & ${}^L$  & ${}^L$ IIexp\\
$33.4$ & $30.9$ & $31.9$ $R=40kpc$\\
Y & Y & N\\
$3.4$ & $3.15$ & $3.25$ \\
$3.16$ & $2.91$ & $3.01$ \\
\end{tabular}
&
\begin{tabular}{c|c|c}
${}^R$ & ${}^M$ & ${}^P$\\
$7.14$ & $8.55$ & $11.5$ \\
N & Y & Y\\
$4.35$ & $5.21$ & $7.01$ \\
$3.49$ & $4.35$ & $6.15$ \\
\end{tabular}
\\
\end{tabular}
\begin{tabular}{r|c|c}
\hline
name     &
NGC 4736 &
UGC 6446 \\
\hline
incl. angle [deg] &
$35$  ${}^B$         &
$44$ ${}^D$\\
distance [Mpc] &
$4.7$ ${}^P$ &
$18.6$ ${}^G$\\
morphological type &
Sab ${}^B$ &
Sd ${}^G$ \\
$L_{B}$ $\sqq{10^{10}L_{\odot}}$ &
$1.3$ ${}^S$ &
$0.218$ ${}^J$\\
$M_{H+He}$ $\sqq{10^{10}M_{\odot}}$ &
$0.067$ ${}^T$ &
$0.434$ ${}^G$\\
$M_{H_2}$ $\sqq{10^{10}M_{\odot}}$ &
- &
- \\
\begin{tabular}{r}
rotation curve \\
$M_{Gal}$ $\sqq{10^{10}M_{\odot}}$ \\
breaking sphericity test\\
${M_{Gal}}/{L_{B}}$\\
$\frac{M_{Gal}-M_{gaz}}{L_{B}}$
\end{tabular}
&
\begin{tabular}{r}
Thing \\
$2.68$ \\
Y\\
$2.06$\\
$2.01$
\end{tabular}
& \begin{tabular}{r}
${}^G$ \\
1.5 \\
N\\
6.88\\
4.89
\end{tabular}
\\
\end{tabular}
\caption{\label{tab:2}Results obtained in the framework of global disk model for
various spiral galaxies and references to the used measurement data: (IIexp=double exponential falloff), ${}^A$ -- \citep{bib:carignanpuche}, ${}^B$ -- \citep{bib:sofue},
${}^D$ -- \citep{bib:20}, ${}^E$ -- \citep{bib:13}, ${}^F$ -- \citep{bib:23}, ${}^G$ -- \citep{bib:19}, ${}^H$ -- \citep{bib:3}, ${}^I$ -- \citep{bib:14}, ${}^J$ -- \citep{bib:18}, ${}^K$ -- \citep{bib:11}, ${}^L$ -- \citep{bib:zanmar}, ${}^M$ -- \citep{bib:boomsma},  ${}^O$ -- \citep{bib:dicaire}, ${}^P$ -- \citep{bib:blok}, ${}^R$ -- \citep{bib:sanders}, ${}^S$ -- \citep{bib:munoz}, ${}^T$ -- \citep{bib:mulder}
}
\end{minipage}
\end{table*}

\section*{Acknowledgments}
We acknowledge the usage of the HyperLeda database (http://leda.univ-lyon1.fr)

\appendix
\section{Appendix}
\begin{figure}
\includegraphics[width=\columnwidth,height=0.8\columnwidth]{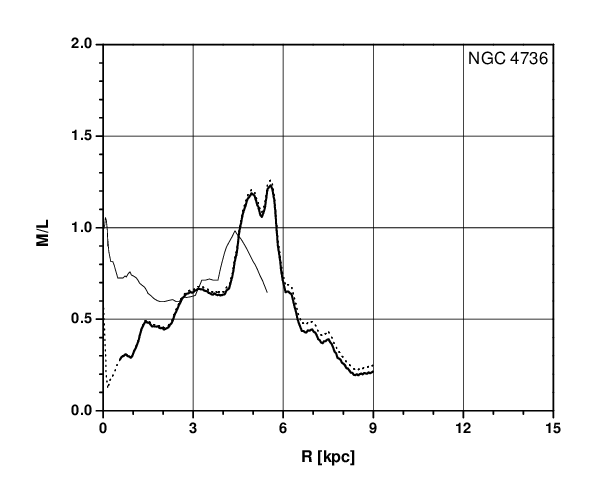}
\includegraphics[width=\columnwidth,height=0.8\columnwidth]{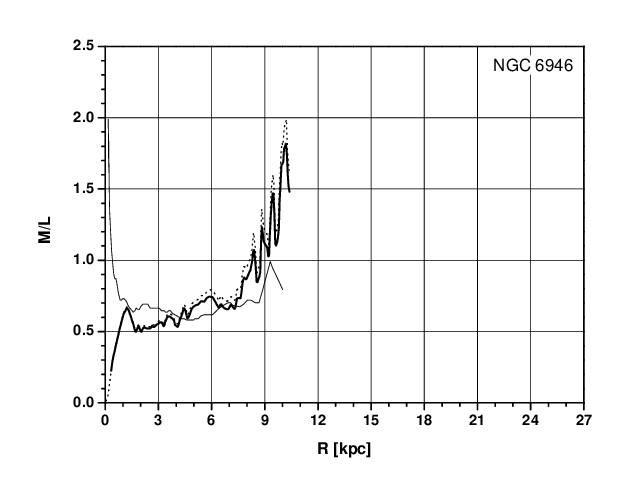}
\includegraphics[width=\columnwidth,height=0.8\columnwidth]{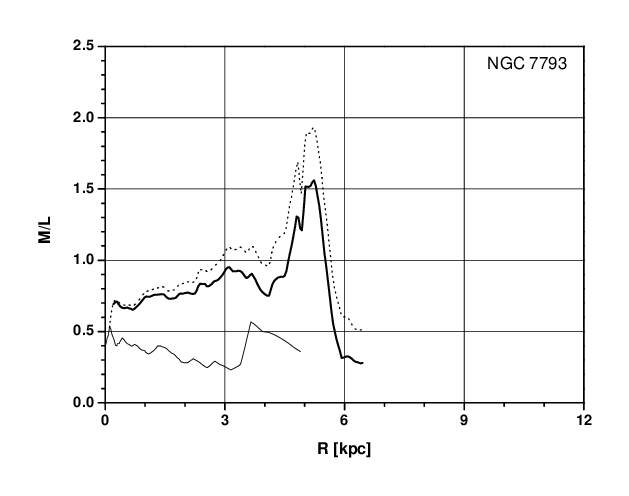}
\caption{\label{fig:appendix} Comparison of mass-to-light ratio profiles for galaxies NGC 4736, NGC 6946 and NGC 7793.
[dotted line] -- global disk model mass-to-light ratio profiles derived based on THINGS rotation curves and $3.6\mathrm{\mu{}m}$ brightness profiles from the Spitzer observations \citep{bib:blok}; [thick solid line] the same as before but with HI and He mass contribution subtracted; and for comparison -- [thin solid line]  -- mass-to-light ratio profiles determined in \citep{bib:blok} based on J-K colors.
}\end{figure}

Let us compare our results concerning mass-to-light ratios -- obtained based on THINGS rotation curves and $3.6\mathrm{\mu{}m}$ brightness profiles\footnote{Corrected for averaged inclination given in \citep{bib:blok}} from the Spitzer observations \citep{bib:blok} -- with (stellar) mass-to-light ratios determined in \citep{bib:blok} based on the observed J--K colours. It should be stressed that our mass-to-light ratios are dynamical; that is, they are calculated as a ratio of the local value of surface mass density inferred from galaxy rotation to the local value of surface brightness. To obtain stellar mass-to-light ratio profiles, one should first subtract the gas contribution from the total mass density (assuming that only baryonic matter is present). The result of this comparison is shown in Fig.  \ref{fig:appendix}.
It is seen that in the case of galaxy NGC 6946 there is no significant difference between the two results. The only discrepancy is in the central galaxy region. However, the THINGS rotation curve for this galaxy was measured starting from radius $r=1.38\kpc$, consequently, it was not possible using these incomplete data to determine the mass density in this region by direct use of equation  \eqref{eq:sigmamoja}, Therefore the mass-to-light ratio profile could not be well determined in this region either. There is a larger discrepancy in the case of galaxy NGC 7793. Our mass-to-light ratio profile is everywhere higher than that in \citep{bib:blok}.
Concerning galaxy NGC 4736, our results agree with \citep{bib:blok} in the region beyond radius $2\kpc$.

There are two reasons for these differences. The first is the inaccuracy in the method used in \citep{bib:blok} to determine mass-to-light ratios. The second is the fact that the ratios in \citep{bib:blok} are those of stars alone. Our mass-to-light ratios include other mass components. We subtracted the contribution from neutral hydrogen and helium, but there are other mass components present in each galaxy, such as dust or neutron stars, that modify this ratio. It should be noted that in the case of the three galaxies the average values of the ratios are low, namely $0.4$, $0.88$ and $1.02$, for NGC 4736, 6946 and 7793 (gas subtracted), respectively, and these values provide an additional argument against the presence of non-baryonic matter in significant amounts in these galaxies.


\begin{thebibliography}{99}

\bibitem[\protect\citeauthoryear{Athanassoula \& Sellwood}{1987}]{bib:ref1}
Athanassoula \& Sellwood (1987) in Kormendy J., Knapp G. R., eds, Proc. IAU Symp. 117, Dark Matter in the Universe. Reidel, Dordrecht,
p. 300
\bibitem[\protect\citeauthoryear{Athanassoula et al.}{1987}]{bib:ref2} Athanassoula, E., Bosma, A.,  Papaioannou, S. (1987) A\&A, 179, 23
\bibitem[\protect\citeauthoryear{Bahcall \& Casertano}{1985}]{bib:B1}
Bahcall, J. N., \& Casertano (1985) ApJ, 293, L7
\bibitem[\protect\citeauthoryear{Battaner at al.}{1992}]{bib:B10} Battaner, E.; Garrido, J. L.; Membrado, M.; Florido, E (1992) Nature, 360, 6405, p. 652
\bibitem[\protect\citeauthoryear{Bernabei et al.}{2007}]{bib:27} Bernabei R. at al., 2007, Nucl. Phys. B, Proc. Suppl., 166, 87
\bibitem[\protect\citeauthoryear{Binney \& Tremaine}{1987}]{bib:A11}
Binney, J., \& Tremaine, S. (1987), Galactic Dynamics. Princeton
Univ. Press, Princeton
\bibitem[\protect\citeauthoryear{Boomsma et al.}{2008}]{bib:boomsma}
     Boomsma R., T. A. Oosterloo, F. Fraternali, J. M. van der Hulst and R. Sancisi (2008) A\&A, 490, 555
\bibitem[\protect\citeauthoryear{Bosma}{1978}]{bib:C7}
Bosma, A. (1978), Ph.D. thesis, Univ. Groningen
\bibitem[\protect\citeauthoryear{Bosma}{1999}]{bib:C8}
Bosma, A. (1999) in Merritt D. R., Valluri M., Sellwood J. A., eds, ASP
Conf. Ser. Vol. 182, Galaxy Dynamics. Astron. Soc. Pac., San Francisco,
p. 339
\bibitem[\protect\citeauthoryear{Bratek et al.}{2008}]{bib:bratek_MNRAS} Bratek {\L}., Ja{\l}ocha J., Kutschera M., 2008, MNRAS, 391, 1373
\bibitem[\protect\citeauthoryear{Buchhorn}{1992}]{bib:A21} Buchhorn, M. (1992), Ph.D. thesis, Australian Natl. Univ.
\bibitem[\protect\citeauthoryear{Carignan}{1985}]{bib:11} Carignan C., 1985, ApJS, 58, 107
\bibitem[\protect\citeauthoryear{Carignan \& Puche}{1990}]{bib:carignanpuche} Carignan C.,  Puche D., 1990, AJ, 100, 394
\bibitem[\protect\citeauthoryear{Ciardullo, Durrell, at al}{2004}]{bib:A7}
Ciardullo R., Durrell P. R., Laychak M. B., Herrmann K. A., Moody K.,
Jacoby G. H., Feldmeier J. J., (2004) ApJ, 614, 167
\bibitem[\protect\citeauthoryear{Clowe et al.}{2006}]{bib:25} Clowe D., Brada\v{c} M., Gonzalez A.H., Markevitch M., Randall S.W., Jones C., Zaritsky D., 2006, ApJ, 648, L109
\bibitem[\protect\citeauthoryear{de Blok, McGaugh \& Rubin}{2001}]{bib:A8}
de Blok W. J. G., McGaugh S. S., Rubin V. C., 2001, AJ, 122, 2396
\bibitem[\protect\citeauthoryear{de Blok et al.}{2008}]{bib:blok}de BlokW. J. G.,Walter F., Brinks E., Trachternach C., Oh S.-H., Kennicutt
R. C., 2008, AJ, 136, 2648
\bibitem[\protect\citeauthoryear{Dicaire et al.}{2008}]{bib:dicaire}
Dicaire, I.; Carignan, C.; Amram, P.; Marcelin, M.; Hlavacek-Larrondo, J.; de Denus-Baillargeon, M.-M.; Daigle, O.; Hernandez, O. (2008)
AJ, 135, 6, 2038-2047.
\bibitem[\protect\citeauthoryear{Evans}{2001}]{bib:A10}
Evans N. W., 2001, in Spooner N. J. C., Kudryavtsev V., eds, Proc. 3rd
InternationalWorkshop on the Identification of DarkMatter.World Sci.,
Singapore, p. 85
\bibitem[\protect\citeauthoryear{Fall \& Efstathiou}{1980}]{bib:C2}
Fall S. M., Efstathiou G., 1980, MNRAS, 193, 189
\bibitem[\protect\citeauthoryear{Herrmann and Ciardullo}{2005}]{bib:A6}Herrmann K. A., Ciardullo R., 2005, in Szczerba R., Stasinska G., G\'{o}rny
S. K., eds, AIP Conf. Proc. Vol. 804, Planetary Nebulae as Astronomical
Tools. Am. Inst. Phys., New York, p. 341
\bibitem[\protect\citeauthoryear{Hoekstra et al.}{2001}]{bib:23} Hoekstra H., van Albada T.S., Sancisi R., 2001, MNRAS, 323, 453
\bibitem[\protect\citeauthoryear{Hohl}{1976}]{bib:C3}
Hohl F., 1976, AJ, 81, 30
\bibitem[\protect\citeauthoryear{HyperLeda}{}]{bib:20} HyperLeda database (http://leda.univ-lyon1.fr)
\bibitem[\protect\citeauthoryear{Ja{\l}ocha et al.}{2007}]{bib:22} Ja{\l}ocha J., Bratek {\L}., Kutschera M., Kolonko M., 2007,
Acta Phys. Pol. B, 38, 3859
\bibitem[\protect\citeauthoryear{Ja{\l}ocha et al.}{2008}]{bib:jalocha_ApJ} Ja{\l}ocha J., Bratek {\L}., Kutschera M., 2008, ApJ, 679, 373
\bibitem[\protect\citeauthoryear{Jorsater \& Moorsel}{1995}]{bib:13} Jorsater S., van Moorsel G.A., 1995, AJ, 110, 2037
\bibitem[\protect\citeauthoryear{Kalnajs}{1983}]{bib:A19}
Kalnajs A. J., 1983, in Athanassoula E., ed., Proc. IAU Symp. 100, Internal
Kinematics and Dynamics of Galaxies. Reidel, Dordrecht, p. 87
\bibitem[\protect\citeauthoryear{Kalnajs}{1987}]{bib:A5}
Kalnajs (1987) IAUS, 117, 289K
\bibitem[\protect\citeauthoryear{Kent}{1986}]{bib:A20} Kent, S. M. 1986, AJ, 91, 1301
\bibitem[\protect\citeauthoryear{Kent}{1987}]{bib:24} Kent S.M., 1987, AJ, 93, 816
\bibitem[\protect\citeauthoryear{Kutschera \& Ja{\l}ocha}{2004}]{bib:B9} Kutschera, M.; Ja{\l}ocha, J. (2004)
	Acta Phys. Polonica B,  35, 10, 2493
\bibitem[\protect\citeauthoryear{Lake \& Feinswog}{1989}]{bib:B3}
Lake, G., Feinswog, L. (1989) AJ, 98, 166
\bibitem[\protect\citeauthoryear{Maller et al.}{2000}]{bib:C5}Maller A. H., Simard L., Guhathakurta P., Hjorth J., Jaunsen A. O., Flores
R. A., Primack J. R., 2000, ApJ, 533, 194
\bibitem[\protect\citeauthoryear{Mulder \& van Driel}{1993}]{bib:mulder}
Mulder, P. S., and van Driel, W. (1993) A\&A, 272, 63
\bibitem[\protect\citeauthoryear{Mu\~{n}oz-Tu\~{n}\'{o}n et al.}{1989}]{bib:munoz}
   Mu\~{n}oz-Tu\~{n}\'{o}n C., Prieto M., Beckman J. E., Cepa J., 1989, Ap\&SS, 156,
301
\bibitem[\protect\citeauthoryear{Olling}{1996}]{bib:B5}
Olling, R. P. (1996), AJ, 112, 481
\bibitem[\protect\citeauthoryear{Ostriker \& Peebles}{1973}]{bib:A2}
Ostriker J. P., Peebles P. J. E., 1973, ApJ, 186, 467
\bibitem[\protect\citeauthoryear{Palunas \& Williams}{2000}]{bib:A17}
Palunas P., Williams T. B., 2000, AJ, 120, 2884
\bibitem[\protect\citeauthoryear{Rix \& Zaritsky}{1995}]{bib:A23} Rix H.-W., Zaritsky D., 1995, ApJ, 447, 82
\bibitem[\protect\citeauthoryear{Rubin, Thonnard \& Ford}{1978}]{bib:C6} Rubin, V. C., Thonnard, N., \& Ford, W. K. J. (1978)  ApJ, 225, L107
\bibitem[\protect\citeauthoryear{Sackett}{1996}]{bib:A15}
Sackett (1996) IAUS, 173, 165
\bibitem[\protect\citeauthoryear{Sackett}{1997}]{bib:A14}
Sackett (1997) ApJ, 483, 103
\bibitem[\protect\citeauthoryear{Sackett et al.}{1994}]{bib:B4}
Sackett, P. D., Rix, H.-W., Jarvis, B. J. \& Freeman, K. C. (1994), ApJ, 436, 629
\bibitem[\protect\citeauthoryear{S\`{a}nchez-Salcedo \& Reyes-Ruiz}{2004}]{bib:B8}
S\'{a}nchez-Salcedo F. J., Reyes-Ruiz M., 2004, ApJ, 607, 247
\bibitem[\protect\citeauthoryear{Sanders}{1996}]{bib:sanders}  Sanders R. H. (1996) ApJ, 473, 117-129
\bibitem[\protect\citeauthoryear{Sellwood}{1980}]{bib:C1}
Sellwood J. A., 1980, A\&A, 89, 296
\bibitem[\protect\citeauthoryear{Sellwood}{1983}]{bib:A1}
Sellwood J. A., 1983, in Athanassoula E., ed., Proc. IAU Symp. 100, Internal
Kinematics and Dynamics of Galaxies. Reidel, Dordrecht, p. 197
\bibitem[\protect\citeauthoryear{Sellwood}{1998}]{bib:C9}
Sellwood J. A., 1998, in Merritt D. R., Valluri M., Sellwood J. A., eds, ASP
Conf. Ser. Vol. 182, Galaxy Dynamics. Astron. Soc. Pac., San Francisco,
p. 339
\bibitem[\protect\citeauthoryear{Sellwood \& Kosowsky}{2000}]{bib:A9}
Sellwood J. A., Kosowsky A., 2000, in Hibbard J. E., Rupen M. P., van
Gorkom J. H., eds, ASP Conf. Ser. Vol. 240, Gas and Galaxy Evolution.
Astron. Soc. Pac., San Francisco, p. 311
\bibitem[\protect\citeauthoryear{Sellwood \& Evans}{2001}]{bib:B11} Sellwood J. A., Evans N. W., 2001, ApJ, 546, 176
\bibitem[\protect\citeauthoryear{Sofue}{1997}]{bib:sofue} Sofue Y., 1997, PASJ, 49, 17
\bibitem[\protect\citeauthoryear{Sug-Whan \& Muk-Suk}{1984}]{bib:3} Sug-Whan K., Muk-Suk C., 1984, J. Korean Astron. Soc., 17, 23
\bibitem[\protect\citeauthoryear{Toomre}{1964}]{bib:B6}
Toomre A., 1964, ApJ, 139, 1217
\bibitem[\protect\citeauthoryear{Tully et al.}{1996}]{bib:18} Tully R. B., Verheijen M. A. W., Pierce M. J., Huang J.-S., Wainscoat R. J., 1996, AJ, 112, 2471
\bibitem[\protect\citeauthoryear{van Albada \& Sancisi}{1986}]{bib:B2}
van Albada G. D., Sancisi R., 1986, Phil. Trans. R. Soc. Lond. A, 320, 447
\bibitem[\protect\citeauthoryear{Verheijen \& Sancisi}{2001}]{bib:19} Verheijen M. A. W., Sancisi R., 2001, A\&A, 370, 765
\bibitem[\protect\citeauthoryear{Williams et al.}{2009}]{bib:A18}
 Williams M. J., Bureau M., Cappellari M., 2009, MNRAS, 400, 1665
\bibitem[\protect\citeauthoryear{Young \& Scoville}{1991}]{bib:14} Young J.S.,  Scoville N.Z., 1991, ARA\&A, 29, 581
\bibitem[\protect\citeauthoryear{Zanmar et al.}{2008}]{bib:zanmar} Z\'{a}nmar S\'{a}nchez R., Sellwood J. A., Weiner B. J., Williams T. B., 2008,
ApJ, 674, 797
\end{thebibliography}
\end{document}